\documentstyle[12pt]{article} 
\input psfig.sty

\def\Title#1{\begin{center} {\Large #1 } \end{center}}
\def\Author#1{\begin{center}{ \sc #1} \end{center}}
\def\Address#1{\begin{center}{ \it #1} \end{center}}

\def\doeack{\footnote{Work supported by the Department of Energy,
                     contract DE--AC03--76SF00515.}}
\def\SLAC{Stanford Linear Accelerator Center\\
    Stanford University, Stanford, California 94309 USA}

\newenvironment{Abstract}{\begin{quotation} \begin{center}
                       ABSTRACT
     \end{center}\bigskip  }{\end{quotation}}
\def\beq{\begin{equation}}
\def\eeq#1{\label{#1}\end{equation}}
\def\eeqn{\end{equation}}
\def\beqa{\begin{eqnarray}}
\def\eeqa#1{\label{#1}\end{eqnarray}}
\def\eeqan{\end{eqnarray}}

\def\Acknowledgements{\bigskip  \bigskip \begin{center} \begin{large}
             \bf ACKNOWLEDGEMENTS \end{large}\end{center}}

\def\Re{{\cal R \mskip-4mu \lower.1ex \hbox{\it e}\,}}
\def\Im{{\cal I \mskip-5mu \lower.1ex \hbox{\it m}\,}}
\def\nn{\noindent}
\def\ie{{\it i.e.}}
\def\eg{{\it e.g.}}

\def\etal{{\it et al.}}
\def\ibid{{\it ibid}.}
\def\sub#1{_{\lower.25ex\hbox{$\scriptstyle#1$}}}
\def\sul#1{_{\kern-.1em#1}}
\def\sll#1{_{\kern-.2em#1}}  
\def\sbl#1{_{\kern-.1em\lower.25ex\hbox{$\scriptstyle#1$}}}
\def\ssb#1{_{\lower.25ex\hbox{$\scriptscriptstyle#1$}}}
\def\sbb#1{_{\lower.4ex\hbox{$\scriptstyle#1$}}}

\def\to{\rightarrow}
\def\dk{\ifmmode \Delta\kappa\else $\Delta\kappa$\fi}
\def\sigt{\ifmmode \tilde\sigma\else $\tilde\sigma$\fi}
\def\mh{\ifmmode m\sbl H \else $m\sbl H$\fi}
\def\mch{\ifmmode m_{H^\pm} \else $m_{H^\pm}$\fi}
\def\mt{\ifmmode m_t\else $m_t$\fi}
\def\mc{\ifmmode m_c\else $m_c$\fi}
\def\mz{\ifmmode M_Z\else $M_Z$\fi}
\def\mw{\ifmmode M_W\else $M_W$\fi}
\def\mws{\ifmmode M_W^2 \else $M_W^2$\fi}
\def\mhs{\ifmmode m_H^2 \else $m_H^2$\fi}   
\def\mzs{\ifmmode M_Z^2 \else $M_Z^2$\fi}
\def\mts{\ifmmode m_t^2 \else $m_t^2$\fi}
\def\mcs{\ifmmode m_c^2 \else $m_c^2$\fi}
\def\mchs{\ifmmode m_{H^\pm}^2 \else $m_{H^\pm}^2$\fi}
\def\ztwo{\ifmmode Z_2\else $Z_2$\fi}
\def\zone{\ifmmode Z_1\else $Z_1$\fi}
\def\mtwo{\ifmmode M_2\else $M_2$\fi}
\def\mone{\ifmmode M_1\else $M_1$\fi}
\def\tb{\ifmmode \tan\beta \else $\tan\beta$\fi}
\def\xw{\ifmmode x\sub w\else $x\sub w$\fi}
\def\ch{\ifmmode H^\pm \else $H^\pm$\fi}
\def\lum{\ifmmode {\cal L}\else ${\cal L}$\fi}
\def\inpb{\ifmmode {\rm pb}^{-1}\else ${\rm pb}^{-1}$\fi}
\def\infb{\ifmmode {\rm fb}^{-1}\else ${\rm fb}^{-1}$\fi}
\def\epem{\ifmmode e^+e^-\else $e^+e^-$\fi}
\def\ppb{\ifmmode \bar pp\else $\bar pp$\fi}

\def\bsg{\ifmmode b\rightarrow s\gamma \else $b\rightarrow s\gamma$\fi}
 
\newskip\zatskip \zatskip=0pt plus0pt minus0pt
\def\matth{\mathsurround=0pt}

\def\atversim#1#2{\lower0.7ex\vbox{\baselineskip\zatskip\lineskip\zatskip
  \lineskiplimit 0pt\ialign{$\matth#1\hfil##\hfil$\crcr#2\crcr\sim\crcr}}}

\begin{document}
\rightline{\vbox{\halign{&#\hfil\cr
&SLAC-PUB-7153\cr
&May 1996\cr}}}
\vspace{0.8in} 
\Title{Probing Weak Anomalous Top Quark Couplings with Final State Gluons 
at the NLC 
}
\bigskip
\Author{Thomas G. Rizzo\doeack}
\Address{\SLAC}
\bigskip
\begin{Abstract}
 
The rate and corresponding gluon jet energy distribution for the process 
$e^+e^- \to t\bar tg$ are sensitive to the presence of anomalous dipole-like 
couplings of the top to the photon and $Z$ at the production vertex. For 
sizeable anomalous couplings of this type substantial deviations from the 
expectations of the Standard Model are likely. We explore the capability of 
the NLC to discover or place bounds on these types of top quark couplings. 
The resulting constraints are found to be quite complementary to those which 
arise from direct probes of the top quark production vertex. 

\end{Abstract}
\bigskip

\vskip1.0in
\begin{center}
To appear in {\it Physics and Technology of the Next Linear 
Collider}, eds. D.\ Burke and M.\ Peskin, reports submitted to Snowmass 1996.
\end{center}
\bigskip

\def\thefootnote{\fnsymbol{footnote}}
\setcounter{footnote}{0}
\newpage
\section{Introduction}

The Standard Model(SM) has provided a remarkably successful description of 
almost all available data involving the strong and electroweak interactions. 
In particular, the discovery of the top quark at the Tevatron with a 
mass{\cite {tev}}, $m_t=175\pm 9$ GeV, close to that anticipated by fits to 
precision electroweak data is indeed a great triumph. However, 
the fact that $R_b$(and perhaps $A_b$) remains{\cite {moriond}} more than 
$3.3(1.8)\sigma$ from SM expectations 
may be providing us with the first indirect window 
into new physics. In fact, this apparent deviation in $b$-quark couplings 
from the SM expectations could indicate that some new physics is interacting 
with the third family as 
a whole. Since the top is the most massive fermion, it is believed by many 
that the detailed physics of the top quark may be significantly different 
than what is predicted by the SM. This suggestion makes precision measurements 
of all of the top quark's properties mandatory. 

Perhaps the most obvious and easily imagined scenario is one in which 
the top's 
couplings to the SM gauge bosons, \ie, the $W$, $Z$, $\gamma$, and $g$, 
are altered. This possibility, extended to all of the fermions of the third 
generation, has attracted a lot of attention over the last few 
years{\cite {big}}. In the case of the electroweak interactions involved in 
top pair production in $e^+e^-$ collisions, the lowest dimensional 
gauge-invariant operators representing new physics that we can introduce take 
the form of dipole moment-type couplings to the $\gamma$ and $Z$. The 
anomalous magnetic moment-type operators, 
which we can parameterize by a pair of dimensionless quantities, 
$\kappa_{\gamma,Z}$, are $CP$-conserving. The corresponding electric dipole 
moment terms, parameterized as $\tilde \kappa_{\gamma,Z}$, are $CP$-violating. 
The shift in the three-point $t\bar t\gamma$ and $t\bar tZ$ interactions due 
to the existence of these anomalous couplings can be written as 
\begin{equation}
{\delta \cal L}={i\over {2m_t}}\bar t
\sigma_{\mu\nu}q^{\nu}\left[e(\kappa_\gamma^t-i\tilde\kappa_
\gamma^t\gamma_5)A^{\mu}+{g\over {2c_w}}(\kappa_Z^t-i\tilde\kappa_Z^t
\gamma_5)Z^{\mu}\right]t \,,
\end{equation}
where $e$ is the proton charge, $g$ is the standard weak coupling constant, 
$c_w=cos \theta_W$, and $q$ is the $\gamma$ or $Z$'s four-momentum. Gauge 
invariance will also lead to new four-point interactions involving two  
gauge bosons and the top, \eg, $t\bar t \gamma\gamma, ZZ, Z\gamma, W^+W^-$,  
but they will not concern us here as we will only work to leading order in the 
electroweak interactions. In most cases gauge invariance will 
relate any $t\bar tZ,\gamma$ anomalous couplings to others involving the 
$tbW$ vertex. Escribano and Masso{\cite {big}} have shown that in general all 
of the anomalous three-point couplings involving the neutral gauge bosons can  
be unrelated even when the underlying operators are SM gauge invariant. Thus in 
our analysis we will treat $\kappa_{\gamma,Z}^t$ and 
$\tilde \kappa_{\gamma,Z}^t$ as independent free parameters. (Of course, 
within any  
particular new physics scenario the anomalous couplings will no longer be 
independent.) As has been discussed in the literature{\cite {big}}, if any of 
the anomalous couplings are sufficiently large their effects can be directly 
probed by top pair production. The purpose of the present work is to consider 
the sensitivity of the process $e^+e^- \to t\bar tg$ to non-zero values of the 
$t\bar tZ,\gamma$ anomalous couplings.

\section{Analysis}

In the present analysis we consider how the `normalized' gluon energy 
distribution,   
\begin{equation}
{dR\over {dz}}= {1\over {\sigma(e^+e^-\to t\bar t)}}
{d\sigma(e^+e^-\to t\bar tg)\over {dz}} \,,
\end{equation}
where $z=2E_g/\sqrt s$, can be used to constrain anomalous top couplings to 
the $\gamma$ and $Z$. (We work only to lowest order in $\alpha_s$.) Note that 
the anomalous couplings will contribute to {\it both} the numerator and 
denominator of the expression of $dR/dz$. This implies that the sensitivity 
of $R$ to very large values(with magnitudes $\geq$ 1) of the anomalous 
couplings is quite small. However, for the range of anomalous couplings of 
interest to us significant sensitivity is achieved. We follow the procedure 
given in Ref.{\cite {old}} which also supplies the complete formulae 
for evaluating this gluon energy distribution. 

In comparison to Ref.{\cite{old}}, the present analysis has been extended 
in two ways. ($i$) We 
allow for the possibility that two of the four anomalous couplings may be 
simultaneously non-zero. ($ii$) We lower the cut placed 
on the minimum gluon jet energy, $E_g^{min}$, in performing energy spectrum 
fits. The reasons 
for employing such a cut are two-fold. First, a minimum gluon 
energy is required to identify the event as $t\bar tg$ and not just $t\bar t$. 
The cross section for $e^+e^-\to t\bar tg$ itself is infra-red singular 
though free of co-linear singularities due to the finite top quark mass. 
Second, since the top decays rather quickly, 
$\Gamma_t\simeq 1.45$ GeV for $m_t=175$ GeV, we need to worry about 
`contamination' from the additional gluon radiation generated 
off of the $b$-quarks in the final state subsequent to top decay. 
Such events can be effectively 
removed from our sample if we require that $E_g^{min}/\Gamma_t>>1$. In our 
past analysis we were very conservative in our choices for $E_g^{min}$ in 
order to make 
this ratio as large as possible, \ie, we assumed $E_g^{min}=37.5(200)$ GeV for 
a 500(1000) GeV NLC. It is now believed that 
we can with reasonable justification soften these cuts at least as far as 
25(50) GeV for the same center of mass energies{\cite {orr}}, with a possible 
further softening of the cut at the higher energy machine not unlikely. 
Due to the dramatic  
infra-red behaviour of the cross section, this change in the cuts leads not 
only to an increased statistical power, since more events are included in the 
fit, but also to a longer lever arm to 
probe those events with very large gluon jet energies which have the most 
sensitivity to the presence of anomalous couplings. As one might expect, 
we find constraints on the anomalous couplings which are somewhat stronger 
than what was obtained in our previous analysis{\cite {old}}. 

\vspace*{-0.5cm}
\nn
\begin{figure}[htbp]
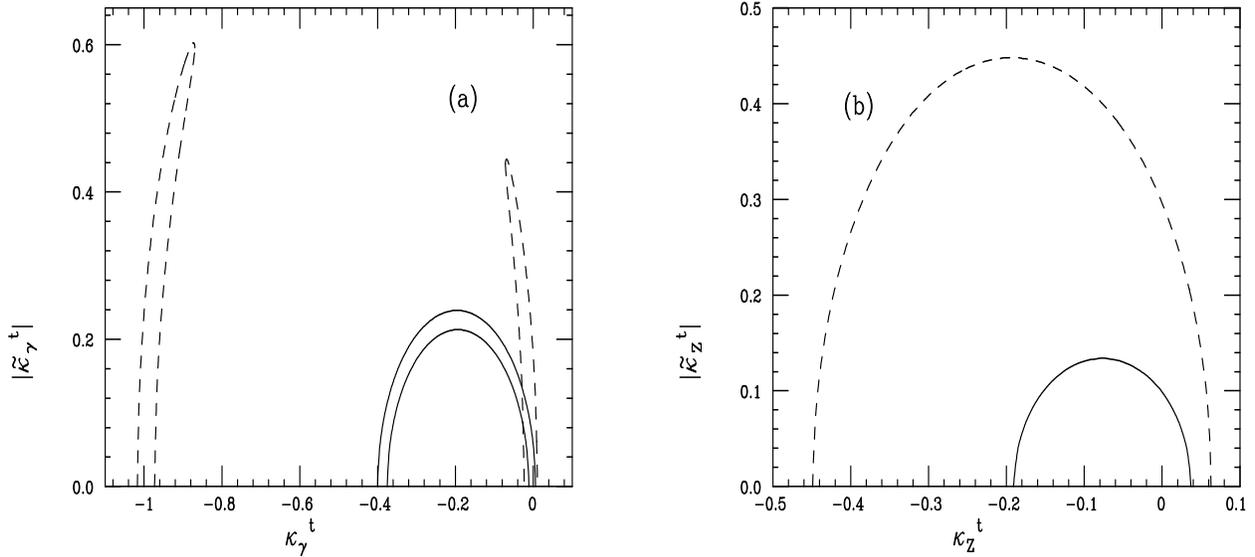

\centerline{
\psfig{figure=ttnewf.res1ps,height=9.1cm,width=9.1cm,angle=-90}
\hspace*{-5mm}
\psfig{figure=ttnewf.res2ps,height=9.1cm,width=9.1cm,angle=-90}}
\vspace*{-1cm}
\caption{\small $95\%$ CL allowed regions obtained for the anomalous couplings 
at a 500(1000) GeV NLC assuming a luminosity of 50(100) $fb^{-1}$ 
lie within the dashed(solid) curves. The gluon energy range $z\geq 0.1$ was 
used in the fit. Only two anomalous couplings are allowed 
to be non-zero at a time.}
\end{figure}
\vspace*{0.4mm}
\vspace*{-0.5cm}
\nn
\begin{figure}[htbp]
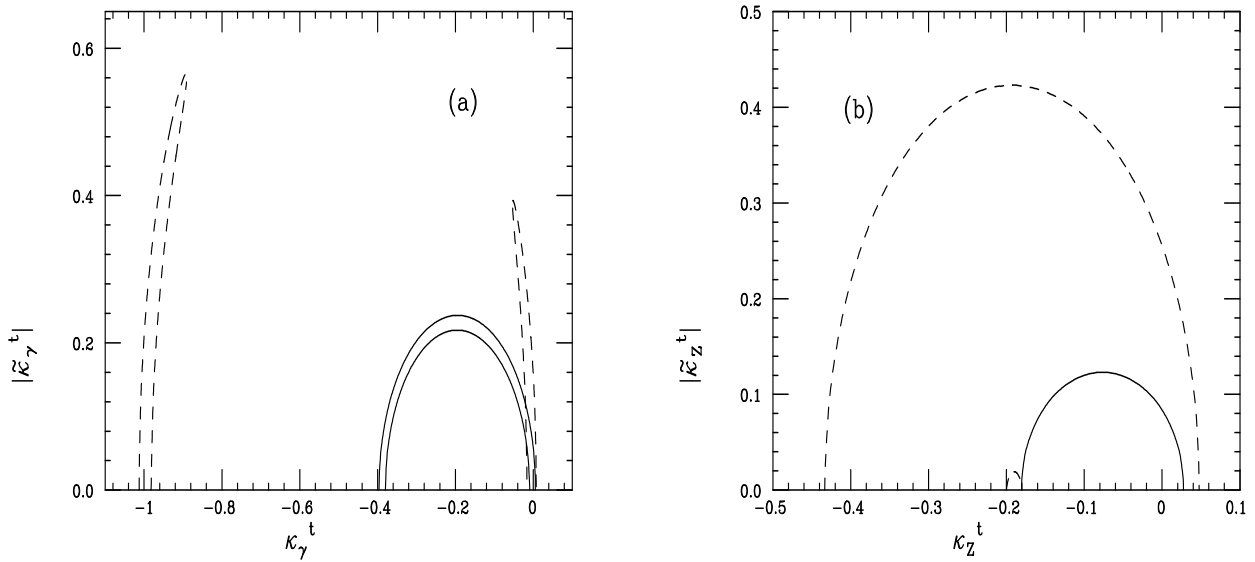

\centerline{
\psfig{figure=ttnewf.res5ps,height=9.1cm,width=9.1cm,angle=-90}
\hspace*{-5mm}
\psfig{figure=ttnewf.res8ps,height=9.1cm,width=9.1cm,angle=-90}}
\vspace*{-1cm}
\caption{\small Same as Fig. 1 but now doubling the integrated luminosity to 
100(200) $fb^{-1}$ for the 500(1000) GeV NLC.}
\end{figure}
\vspace*{0.4mm}

As in our previous work this analysis is based on a Monte Carlo approach. We 
generate data simulating the scaled gluon energy spectrum, $dR/dz$, in 
fixed energy bins 
accounting for only the statistical errors assuming the SM to be correct. A fit 
is then performed to these generated data samples allowing the values of the 
various anomalous couplings to float. In general, one can perform a four 
parameter fit allowing all of the parameters $\kappa_{\gamma,Z}^t$ and 
$\tilde \kappa_{\gamma,Z}^t$ to 
be simultaneously non-zero. Here, for simplicity we allow only two of these 
couplings to be simultaneously non-vanishing, \ie, we consider anomalous top 
couplings to the photon and $Z$ separately. The first results of this analysis 
are shown in Figs.1a-b and Figs. 2a-b which compare a 500 
and 1000 GeV NLC at two different 
values of the integrated luminosity. For the 500(1000) GeV case, the minimum 
gluon jet energy, $E_g^{min}$, was set to 25(50) GeV corresponding to 
$z\geq 0.1$. A fixed energy bin width of $\Delta z$=0.05 was chosen in both 
cases so that at 500(1000) GeV 8(15) bins were used to cover the entire 
spectrum. In either case the constraints on the anomalous $t\bar t\gamma$ 
couplings are seen to be stronger qualitatively than the corresponding 
$t\bar tZ$ ones. In 
the former case the allowed region is essentially a long narrow circular band, 
which has been cut off at the top for the $\sqrt s=$500 GeV NLC. In the 
latter case, the allowed 
region lies inside a rather large ellipse. We see from these figures that to 
increase the sensitivity to anomalous couplings it is far better to go to 
higher center of mass 
energies than to simply double the statistics of the sample. The 1 TeV results 
are seen to be significantly better than those quoted in Ref.{\cite {old}} due 
to lower value of $E_g^{min}$. 

\vspace*{-0.5cm}
\nn
\begin{figure}[htbp]
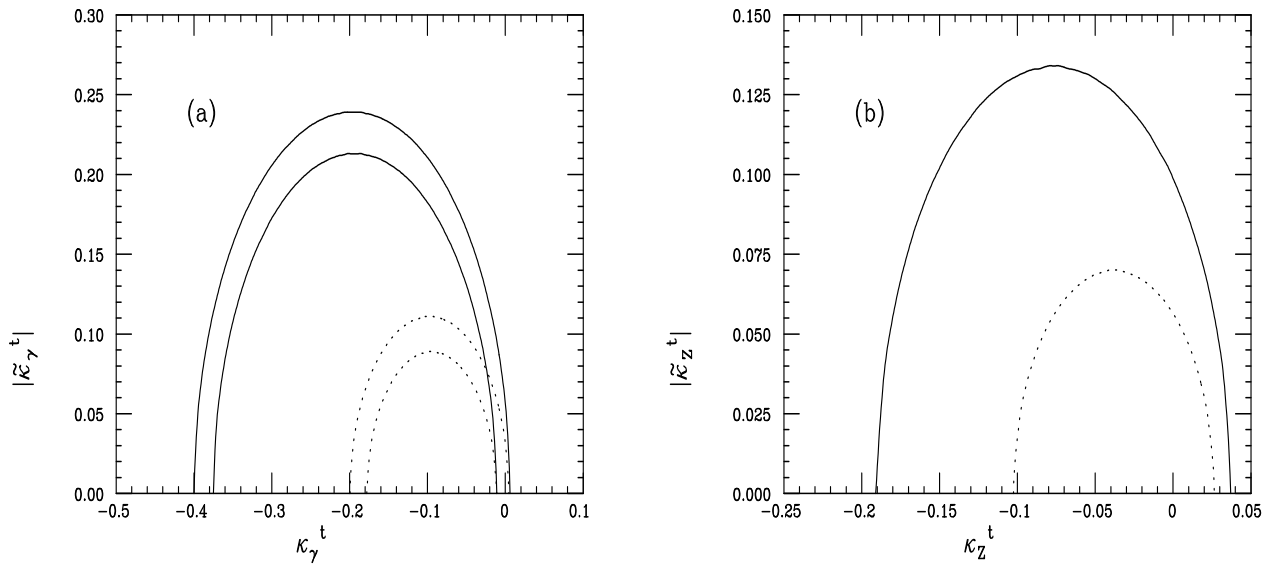

\centerline{
\psfig{figure=ttnewf.res3ps,height=9.1cm,width=9.1cm,angle=-90}
\hspace*{-5mm}
\psfig{figure=ttnewf.res4ps,height=9.1cm,width=9.1cm,angle=-90}}
\vspace*{-1cm}
\caption{\small Same as Fig. 1 but now for a 1 (1.5) TeV NLC assuming an 
integrated luminosity of 100(200) $fb^{-1}$ corresponding to the solid(dotted) 
curve.}
\end{figure}
\vspace*{0.4mm}

A similar pattern is shown in Figs. 3a-b and 4a-b which display and compare 
the result of our fits at center of mass energies of 1 and 1.5 TeV for two 
different integrated luminosities. Note that 
the 1 TeV solid curves shown in Figs. 1 and 2 are expanded by a change of 
scale of roughly a factor of 2 in Figs. 3 and 4. Again we see that the 
anomalous photon 
couplings are far more constrained than are those of the $Z$ and that an 
increase in energy far outweighs additional luminosity if increased 
sensitivity is desired. Fig. 5 shows that for a 1 TeV collider a further 
reduction in $E_g^{min}$ to 25 GeV from 50 GeV does not significantly improve 
our anomalous coupling constraints for either $\gamma$ or $Z$. 

\vspace*{-0.5cm}
\nn
\begin{figure}[htbp]
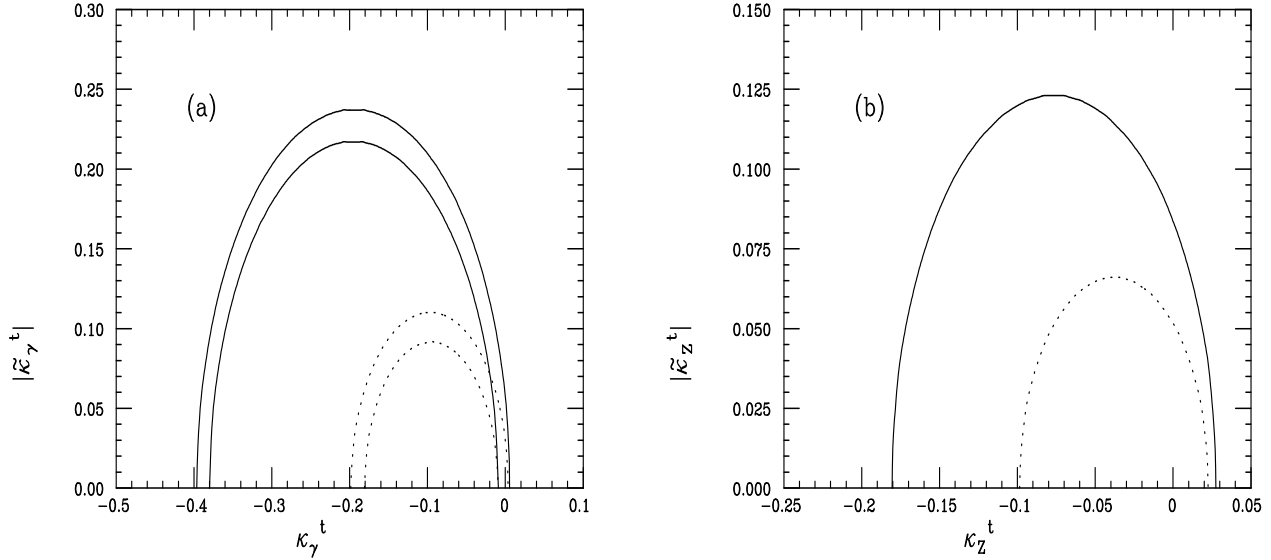

\centerline{
\psfig{figure=ttnewf.res6ps,height=9.1cm,width=9.1cm,angle=-90}
\hspace*{-5mm}
\psfig{figure=ttnewf.res7ps,height=9.1cm,width=9.1cm,angle=-90}}
\vspace*{-1cm}
\caption{\small Same as Fig. 3 but now for luminosities of 200(300) $fb^{-1}$ 
at a 1(1.5) TeV collider corresponding to the solid(dotted) curve.}
\end{figure}
\vspace*{0.4mm}

\section{Discussion and Conclusions}

In this report we have shown that the 
process $e^+e^- \to t\bar tg$ can be used to obtain stringent limits on the 
anomalous dipole-like couplings of the top to both $\gamma$ and $Z$ through 
an examination of the associated gluon energy spectrum. Such 
measurements are seen to be complementary to those which directly probe the 
$t\bar t$ production vertex. By combining both sets of data a very high 
sensitivity to the anomalous couplings can be achieved.

\Acknowledgements

The author would like to thank P. Burrows, A. Kagan and J.L. Hewett for 
discussions related to this work.

\vspace*{-0.5cm}
\nn
\begin{figure}[htbp]
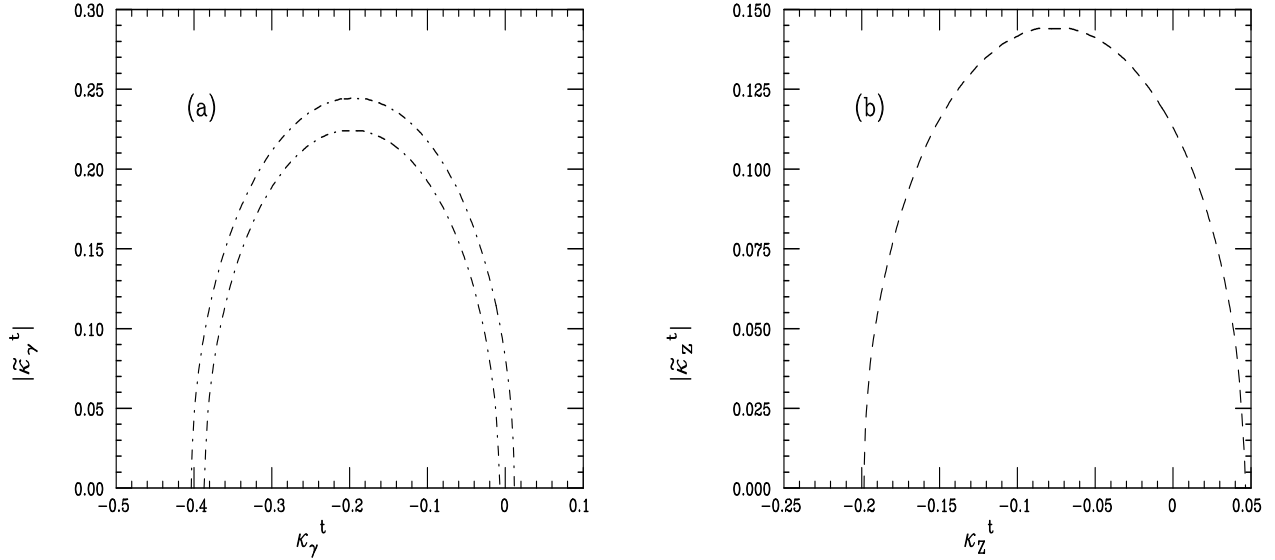

\centerline{
\psfig{figure=ttnewf.res9ps,height=9.1cm,width=9.1cm,angle=-90}
\hspace*{-5mm}
\psfig{figure=ttnewf.res10ps,height=9.1cm,width=9.1cm,angle=-90}}
\vspace*{-1cm}
\caption{\small Same as Fig. 3 for the 1 TeV NLC with a 100 $fb^{-1}$ of 
integrated luminosity but now with a gluon jet cut of $z\geq 0.05$.}
\end{figure}
\vspace*{0.4mm}
%

%
\def\MPL #1 #2 #3 {Mod.~Phys.~Lett.~{\bf#1},\ #2 (#3)}
\def\NPB #1 #2 #3 {Nucl.~Phys.~{\bf#1},\ #2 (#3)}
\def\PLB #1 #2 #3 {Phys.~Lett.~{\bf#1},\ #2 (#3)}
\def\PR #1 #2 #3 {Phys.~Rep.~{\bf#1},\ #2 (#3)}
\def\PRD #1 #2 #3 {Phys.~Rev.~{\bf#1},\ #2 (#3)}
\def\PRL #1 #2 #3 {Phys.~Rev.~Lett.~{\bf#1},\ #2 (#3)}
\def\RMP #1 #2 #3 {Rev.~Mod.~Phys.~{\bf#1},\ #2 (#3)}
\def\ZP #1 #2 #3 {Z.~Phys.~{\bf#1},\ #2 (#3)}
\def\IJMP #1 #2 #3 {Int.~J.~Mod.~Phys.~{\bf#1},\ #2 (#3)}

\end{document}